\documentclass[twocolumn,showpacs,preprintnumbers,amsmath,amssymb]{revtex4}

\usepackage{graphicx}
\usepackage{dcolumn}
\usepackage{bm}

\begin{document}


\title{Quantum squeezing and entanglement in a two-mode Bose-Einstein condensate with 
time-dependent Josephson-like coupling}
\author{S. Choi and N. P. Bigelow}
\affiliation{Department of Physics and Astronomy, University of Rochester, Rochester, New York 14627, USA}

\begin{abstract}
Dynamical evolution of quantum mechanical squeezing and entanglement in a two-mode Bose-Einstein condensate (TBEC) with an adiabatic, time-varying Raman coupling is studied by finding analytical expressions for these quantities. In particular, we study the entanglement between the atoms in the condensate as well as that between the two modes. The nature of the enhanced quantum correlations in TBEC is clarified by considering squeezing and entanglement both with and without the nonlinear interaction turned on; it is found that entanglement approaching maximal value can be achieved with the nonlinear interactions present. Somewhat counter-intuitively, greater squeezing is found in the absence of nonlinear interactions. This is due to the collapses and revivals of the TBEC quantum state induced by the nonlinear interactions. In addition, results involving the self-trapping phase state of TBEC indicate potential for creating a dynamically stable, macroscopic entangled quantum state which is relatively robust with respect to atom number fluctuations.
\end{abstract}


\maketitle

\section{Introduction}

Bose-Einstein condensates (BECs) provide a useful system for investigating matter wave quantum squeezing and entanglement. Squeezing in BECs has already been investigated previously from
a number of different perspectives\cite{squeeze1,squeeze2,squeeze3,squeeze4,squeeze5}.
Squeezed states are quantum states for which no classical analog exists\cite{WallsNature}. The definition of quantum mechanical squeezing, or the reduction of quantum
fluctuations below the standard quantum limit (SQL), derives directly from the Heisenberg Uncertainty Principle:
For two arbitrary operators $\hat{A}$ and $\hat{B}$ which obey the
commutation relation $[\hat{A},\hat{B}]=\hat{C}$, quantum squeezing exists
when one of the variables satisfies the relation\cite{wallszoller}
\begin{equation}
\lbrack \Delta \hat{A}]^{2}<\frac{1}{2}|\langle \hat{C}\rangle |.
\label{inequality}
\end{equation}
Although the origin of quantum squeezing in many experiments is enhanced quantum correlations, it must be noted that the inequality (\ref{inequality}) is {\it not} a sufficient proof of enhanced quantum correlations between the particles, as discussed by Kitagawa and Ueda in the context of spin squeezing\cite{ueda}.  A measure of the amount of spin squeezing was  defined as\cite{ueda,wineland}
\begin{equation}
\xi =\frac{2[\Delta J_{\bot }]^{2}}{J},  \label{KUxi}
\end{equation}
where $[\Delta J_{\bot }]^{2}$ is the variance in the direction orthogonal
to the total spin vector for an $N$ spin-$\frac{1}{2}$ particle system. This definition takes into account the effect of quantum
correlations as it derives from the fact that a minimum uncertainty state for $N$ elementary spin-$\frac{1}{2}$ particles has spin $J/2$ ($J =  N/2$) equally distributed over any two orthogonal components normal to the mean spin direction. This follows from the fact that, in the absence of quantum correlations, the total variance in the normal direction is simply given by the sum of the variances of the individual elementary spins. Any state with uncertainty less than the minimum uncertainty state, i.e. spin squeezed due to enhanced quantum correlations between the particles, therefore gives $\xi <1$.  

Quantum entanglement, which is closely related to quantum squeezing, has recently generated a lot of activity amongst researchers, owing to its significant role in the studies of quantum information theory. A large number of theoretical studies exist on generating entangled states using BECs in a variety of different physical settings. Many of these involve two-mode BECs (TBECs) in which atoms in two different hyperfine states are entangled\cite{twocomp1,Micheli,kennedy,twocomp2,twocomp3}. 
In simple terms, a quantum state is said to be entangled when the state cannot be written as a simple product state, the most prominent example being the famous Bell-state for a two particle system: $|\psi \rangle = (|1 \rangle| 0 \rangle + |0 \rangle|1 \rangle)/\sqrt{2}$, where ``1'' and ``0'' represent spin ``up'' and ``down'' states respectively. From the inseparability criterion for the $N$-particle density matrix,  a parameter for the entanglement between atoms in TBECs has been derived\cite{twocomp1,Micheli}, which has a form similar to Eq. (\ref{KUxi}):
\begin{equation}
\xi _{{\bf n}}^{2}=\frac{N[\Delta {\bf n}_{1}\cdot \hat{J}]^{2}}{\langle
{\bf n}_{2}\cdot \hat{J}\rangle ^{2}+\langle {\bf n}_{3}\cdot \hat{J}\rangle
^{2}}  \label{Sxi}
\end{equation}
where ${\bf n}_{i}$, $i=1,2,3$ are unit orthogonal vectors, and $\hat{J}$ is
the well-known Schwinger angular momentum operator representing a many particle system.
It is easy to see that Eq. (\ref{Sxi}) may be viewed as a generalization of Eq. (%
\ref{KUxi}).

On the other hand, it has also been argued that quantum entanglement in a TBEC is more meaningful  when one considers the system as a bipartite system of two modes, analogous to considering entanglement of electromagnetic modes, as opposed to the photons themselves, as there is no definitive measure for entanglement between three or more subsystems\cite{Milburn}. In addition, the two modes of a TBEC are clearly distinguishable and experimentally accessible subsystems. In this case, the standard measure of entanglement is the von Neumann entropy of the reduced density operator of either of the subsystems:
\begin{equation}
E(\rho_A) = - {\rm Tr}  \rho_A \log_2 (\rho_A),  
\end{equation}
where $\rho_A$ is the reduced density operator for subsystem $A$ defined by the partial trace over the other subsystem $B$,  $\rho_A = {\rm Tr}_B \rho$. It can be easily shown that for a general bipartite system of TBECs written in the Fock basis $|\psi (t) \rangle =  \sum_{m} c_{m} (t) |N - m \rangle_A |m \rangle_B$,  $\rho_A = \sum_{m = 0}^{N} |c_m (t)|^2 | m \rangle \langle m |$ and hence the dynamically evolving bipartite entanglement in TBECs can be parameterized by
\begin{equation}
E(t) = - \sum_{m = 0}^{N} |c_m (t)|^2 \log_2 |c_m (t)|^2.  \label{vonNeumann}
\end{equation}

In this paper, we calculate analytically dynamical evolution of quantum squeezing as well as quantum entanglement for both types of system decompositions of a TBEC: $N$ particle subsystems based on Eq. (\ref{Sxi}) and the subsystems composed of the two modes based on Eq. (\ref{vonNeumann}). We shall consider a TBEC with a continuous time-dependent Raman coupling between the two states, with and without nonlinear interactions.  A physical realization of such a system already exists; the Josephson-like coupling between the two
levels is provided by an external laser in the Rb TBEC of Ref. \cite{myatt},
while a state-dependent magnetic field gradient may be applied to induce
Josephson tunneling in the Na spinor system of Ref. \cite{stenger}. We
consider a system with an adiabatically time-varying coupling with an
off-resonance atom-light interaction.  The TBEC is itself an interesting and important system to study: not only does it display effects similar to that of Josephson junction in superconductors, such as the collapse and revival phenomenon in macroscopic scale, it is also the simplest Bose-Hubbard model with a two site lattice potential for which complete description of entanglement can be given.

The central result of this paper is that we use an exact solution to the time-dependent Schr\"{o}dinger equation
to find analytical expressions for the dynamical evolution of squeezing and entanglement.  In particular, we explicitly identify the effect of the nonlinear interactions on squeezing and entanglement. Since one of the comments of Ref. \cite{Milburn} with regard to the atom-atom entanglement parameterized by Eq. (\ref{Sxi}) was that of practicality, we shall restrict our calculations to two different possible initial states for a TBEC, the Dicke state and the phase state. We consider atom-atom entanglement for which clear physical measurement is in principle possible, namely those involving the measurement of the mean and variance of the atom number difference and the relative phase between the two species.

The paper is organized as follows: In Sec. II, we describe, using the Schwinger notation for the angular momentum operators, the TBEC and the
exact solution to the time-dependent Schr\"{o}dinger Equation. In Sec. III,
we present our main results: the dynamics of the TBEC system with a time-dependent Josephson-like coupling between the two species and the evolution of quantum squeezing and entanglement for the two types of subsystem decompositions for various values of laser couplings, with and without the nonlinear interaction. We conclude in Sec. IV.

\section{Formalism}

\subsection{Hamiltonian}

We consider atomic BECs in two different hyperfine states trapped in a single trap,
with a time-varying Raman coupling between the two levels given by a
spatially uniform electromagnetic field. The Hamiltonian for this system, with the
annihilation operators for the two distinct states denoted $\hat{a}$ and $%
\hat{b}$ under the two mode approximation is the following:
\begin{eqnarray}
\hat{H} & = & \hat{H}_a + \hat{H}_b + \hat{H}_{{\rm int}} + \hat{H}_{{\rm las%
}} \\
\hat{H}_{a} & = & \omega_{a}\hat{a}^{\dagger} \hat{a} + \frac{U_{a}}{2} \hat{%
a}^{\dagger}\hat{a}^{\dagger}\hat{a}\hat{a} \\
\hat{H}_{b} & = & \omega_{b}\hat{b}^{\dagger}\hat{b} + \frac{U_{b}}{2} \hat{b%
}^{\dagger}\hat{b}^{\dagger}\hat{b}\hat{b} \\
\hat{H}_{{\rm int}} & = & \frac{U_{ab}}{2} \hat{a}^{\dagger}\hat{a} \hat{b}%
^{\dagger}\hat{b} \\
\hat{H}_{{\rm las}} & = & \Omega(t) ( \hat{a}^{\dagger}\hat{b}e^{i\varphi
(t)} + \hat{b}^{\dagger}\hat{a} e^{-i\varphi (t)} ).
\end{eqnarray}
$\hat{H}_{a}$ and $\hat{H}_{b}$ describe the two condensates undergoing
self-interaction while $\hat{H}_{{\rm int}}$ and $\hat{H}_{{\rm las}}$
describe the condensates interacting with each other via collisions and
laser-induced interactions respectively. $\hat{H}_{{\rm las}}$ describes a
time-dependent coupling, rendering an overall time-dependent Hamiltonian for
the system. The Hamiltonian may be rewritten by employing the Schwinger
notation for the angular momentum operators, namely, $\hat{J}_x = \frac{1}{2}
( \hat{a}^{\dagger}\hat{b} + \hat{b}^{\dagger}\hat{a})$, $\hat{J}_y = \frac{1%
}{2i} ( \hat{a}^{\dagger}\hat{b} - \hat{b}^{\dagger}\hat{a})$, and $\hat{J}%
_z = \frac{1}{2} ( \hat{a}^{\dagger}\hat{a} - \hat{b}^{\dagger}\hat{b})$
with the Casimir invariant $J^2 = \frac{\hat{N}}{2} (\frac{\hat{N}}{2} + 1)$
where $\hat{N} = \hat{a}^{\dagger} \hat{a} + \hat{b}^{\dagger} \hat{b}$ is
the total number operator and is a conserved quantity. Physically, eigenvalues of
the operator $\hat{J}_z$ represents the difference in the number of atoms in
different hyperfine levels, while $\hat{J}_{x}$ and $\hat{J}_{y}$ takes on
the meaning of relative phase between the two species. The $z$ component angular momentum eigenstates, known also as the Dicke
states\cite{bloch,barnettradmore}, are written as $|j, m \rangle$ where $m = -j, \ldots, j$
with $\hat{J}_z |j, m \rangle = m |j, m \rangle$.  Here, $j =  N/2$ and is the quantum number of
angular momentum.  The raising and
lowering operators are defined in the usual way as $\hat{J}_{\pm} = \hat{J}_{x} \pm i \hat{J}_{y} $ such that $\hat{J}_{\pm }|j,m\rangle =\sqrt{(j\mp m)(j\pm
m+1)}|j,m\pm 1\rangle$.
In terms of the angular momentum operators, the Hamiltonian takes the form\cite{choi,chen}:
\begin{equation}
\hat{H}(t) = \omega_{0} \hat{J}_z + q\hat{J}_z^2 + \Omega(t) \left [ \hat{J}%
_{+}e^{i\varphi (t)} + \hat{J}_{-}e^{-i\varphi (t)} \right ],
\label{Hamiltonian}
\end{equation}
where $\omega_{0} = \omega_{a} - \omega_{b} + (N-1) (U_{a} - U_{b})/2$, $q =
(U_{a} + U_{b} - U_{ab})/2$. It is noted that $U_{a}(U_{b})$ or $U_{ab}$ may, in principle, be tuned via Feshbach resonance through the application of an
external magnetic field\cite{feshbach}; the factors $\omega_{0}$ and $q$ are
consequently adjustable parameters.

\subsection{Solution to the time-dependent Schr\"{o}dinger Equation}

An exact solution to the Schr\"{o}dinger Equation
\begin{equation}
i\hbar \frac{d}{dt} |\psi(t)\rangle = \hat{H}(t) |\psi (t) \rangle
\end{equation}
with the time-dependent Hamiltonian $\hat{H}(t)$ of Eq. (\ref{Hamiltonian})
can be given in terms of a time evolution operator $\hat{U}(t)$, $|\psi(t)
\rangle = \hat{U}(t) |\psi(0) \rangle$ \cite{chen}:
\begin{equation}
U(t) = \hat{R}^{\dagger}(t)e^{-iH^{\prime}(t)}\hat{R}(0),  \label{U_t}
\end{equation}
where $\hat{R}$ is a time dependent unitary transformation defined as
\begin{equation}
\hat{R}(t)= \exp \left [ \frac{\lambda}{2} (\hat{J}_{-} e^{-i \varphi (t)} - \hat{J}_{+}
e^{i \varphi (t)}) \right ],  \label{R_t}
\end{equation}
and
\begin{equation}
\hat{H}^{\prime}(t) = \hat{R}\hat{H}(t)\hat{R}^{\dagger} - i \hat{R} \frac{%
\partial}{\partial t} \hat{R}^{\dagger}. \label{Hprimed}
\end{equation}
The parameter $\lambda$ in Eq. (\ref{R_t}) is an auxiliary parameter which
may be chosen to simplify the transformed Hamiltonian, $\hat{H}^{\prime}$.
It can be shown that $\hat{R}$ generates a gauge transformation under which
the time-dependent Schr\"{o}dinger Equation is covariant\cite{liang}:
\begin{equation}
i\hbar \frac{d}{dt} |\psi^{\prime}(t)\rangle = \hat{H}^{\prime}(t)
|\psi^{\prime}(t) \rangle,
\end{equation}
where the transformed state $|\psi^{\prime}(t) \rangle = \hat{R} |\psi (t)
\rangle$. For our case, $\lambda$ is chosen such that the Hamiltonian is
diagonal in the $\hat{J}_z$ representation. With an additional assumption of
adiabaticity conditions $d \varphi/dt \approx 0$ and $d \lambda /dt \approx
0 $, and also assuming that the two-photon transition terms proportional to $%
\hat{J}_{+}^2e^{i2\varphi(t)}$ and $\hat{J}_{-}^2e^{-i2\varphi(t)}$ can be
neglected \cite{tunneling}, one obtains for the transformed Hamiltonian
\begin{equation}
\hat{H}^{\prime}(t) = \sqrt{\omega_0^2 + 4 \Omega^{2}(t)} \hat{J}_z - \frac{q%
}{2} \hat{J}_{z}^2,  \label{Hamiltonian2}
\end{equation}
where $\lambda$ is chosen such that
\begin{equation}
\tan \lambda = - \frac{2\Omega(t)}{\omega_0}.
\end{equation}
The time evolution of any observable $\hat{A}$ is then given in the Heisenberg
picture as $\langle \hat{A}(t) \rangle \equiv \langle \psi(0) |
U^{\dagger}(t) \hat{A} U(t) | \psi(0) \rangle$, where $U(t)$ is defined in
Eqs. (\ref{U_t}-\ref{Hprimed}) and $| \psi(0) \rangle$ is the initial quantum state of the system. We note here that by ignoring the two-photon transitions for the purpose of obtaining analytical solutions, the effect of ``two-axis counter-twisting''\cite{ueda} is not included in our effective Hamiltonian. For concreteness,  we shall consider in this paper the case $\varphi (t)=\Delta t$ where $\Delta$ gives the detuning of the laser from the $|A\rangle \rightarrow |B\rangle $ transition between the two species. Also, we shall consider the case $U_{a} = U_{b} > U_{ab}$ and identical trapping potentials for the two species, corresponding to the $|F=1, M_{F}  = \pm 1 \rangle$ hyperfine states of Na trapped in an optical dipole trap\cite{twocomp1}. This implies $q > 0$ as well as  $\omega_0 = 0$ i.e.  $\lambda = -\pi/2$ in Eq. (\ref{R_t}). 

\subsection{Initial quantum state of TBEC}

We consider in this paper an $SU(2)$ atomic coherent state or a coherent
spin state (CSS), $| \theta, \phi \rangle$\cite{bloch,barnettradmore} as the
initial quantum state for a TBEC. A CSS  is a minimum uncertainty state that describes a system with well-defined relative phase between the two species, and provides a good description of 
TBECs under suitable experimental conditions\cite{savage}. It is defined
mathematically by applying the rotation operator on the extreme Dicke state $%
|j, j \rangle$ or $|j, -j \rangle$. The definition that we use in this paper
is:
\begin{equation}
|\theta, \phi \rangle = \exp \left [ \frac{\theta}{2} ( \hat{J}_{-}
e^{i\phi} - \hat{J}_{+} e^{-i \phi} ) \right ] |j, j \rangle.
\label{coherent}
\end{equation}
A CSS $| \theta, \phi \rangle$ is therefore an eigenstate of the spin
component in the $(\theta, \phi)$ direction $\hat{J}_{\theta, \phi} \equiv
\hat{J}_{x} \sin \theta \cos \phi + \hat{J}_{y} \sin \theta \sin \phi + \hat{%
J}_{z} \cos \phi$ with eigenvalue $J$ where $\theta$ and $\phi$ denote polar
and azimuthal angles. It can be shown, following from Eq. (\ref{coherent}),
that CSS may be written as a superposition state:
\begin{eqnarray}
|\theta, \phi \rangle  & = & \sum_{m = -j}^{j} {\cal R}_{m}^{j}(\theta, \phi) | j, m \rangle ,  \label{coherent2}
\end{eqnarray}
where 
\begin{eqnarray}
{\cal R}_{m}^{j}(\theta, \phi) & = & (C^{2j}_{j + m})^{1/2} \cos^{j+m} \left ( \frac{\theta}{2} \right )\sin^{j-m} \left ( \frac{\theta }{2} \right )  \nonumber \\ 
& & \times e^{i(j-m)\phi} \label{Rjmtp}
\end{eqnarray}
and $C^{n}_{m}$ denotes the combination, $C^{n}_{m} = n!/[(n-m)!m!]$.  
Further extensive discussions on the
properties of the CSS may be found in Refs. \cite{bloch,barnettradmore}. In this
paper, we shall consider initial CSS's that may be produced experimentally, $|\theta = 0, \phi = 0 \rangle$, and $|\theta = \pi/2, \phi = 0 \rangle$. The CSS $|\theta = 0, \phi = 0 \rangle$ is simply a Dicke state $|j,j \rangle$ in which all atoms are found in one
hyperfine level, and the CSS $|\theta = \pi/2, \phi = 0 \rangle$ is an
eigenstate of $\hat{J}_x$ which is a state with a well-defined phase difference $\phi = 0$, and is a phase state of a two-mode boson system\cite{Castin}.  States with different values of $\theta$, although simple to deal with mathematically, are difficult to produce and observe experimentally. For the same reason, we shall restrict our consideration of squeezing to those in the $x$, $y$, and $z$ directions since the variances in the relative phase and atom number are physical quantities clearly measurable without the ambiguity associated with the measurement of variances in other directions.

\section{Results}

In this section, we present various results for the two initial states for a TBEC, the Dicke state and the phase state. 
The coefficient $q$ provides the strength of the scattering interaction between bosons, and its magnitude in relation to the tunnelling coupling $\Omega$ (as we consider $\omega_0 = 0$ in this paper) is an important parameter in determining distinct coupling regimes\cite{leggett}. Of the three coupling regimes, namely, the Rabi ($q/\Omega \ll 1/N$), the Josephson ($1/N \ll q/\Omega \ll N$) and the Fock ($q/\Omega \gg N$) regimes, it is clear that the regimes of more physical interest are the Rabi and the Josephson regimes, since, in the Fock regime the tunnelling coupling is overwhelmed by the nonlinear interaction term, resulting in no new effects of particular interest. This has been confirmed numerically\cite{tonel}. 

In the Rabi regime, the tunnelling coupling dominates with $q/\Omega$ near zero. We shall therefore consider the limiting case of $q = 0$ i.e. the linear regime which not only illustrates the essential physics of the Rabi regime but also demonstrate clearly the effect of nonlinear interaction on squeezing and entanglement. In addition, we shall consider the lower end of the Josephson regime, as it was found that important features rapidly disappear as one nears the Fock regime. The two couplings considered in this paper are therefore $q = 0$ and $q = q_{j}$ with $q_{j}/\Omega = 3/N$ for the Josephson regime, where we choose $N$ to be 400. We also have an additional parameter in our model not considered in some of the previous work, that of a detuning $\Delta$. As will be shown below, this enters mainly as a phase shift which affects the results in a nontrivial manner. To illustrate the effect of this parameter we shall consider detuning of the same magnitude as the $q$ for the Josephson regime,  ($\Delta = q_{j}$) as well as a relatively large value of detuning ($\Delta/\Omega = 10q_{j}$). In particular, the results in the linear ($q=0$) regime will help us see the role of $\Delta$ with respect to $q$. The result for $\Delta = 0$ can be easily interpolated from these two values of detunings so will not be explicitly presented.

\subsection{Dynamics of the macroscopic spin vector}

We first consider how the quantum state of a TBEC represented by the fictitious spin vector evolves in time 
by calculating the expectation values $\langle \hat{J}_{\alpha}(t) \rangle$ 
where $\alpha$ stands for $x$, $y$, and $z$. This later gives us insight into how the squeezing and entanglement in a TBEC dynamically evolves.
We evaluate the
mean values of spin components based on Eq. (\ref{U_t}),
\begin{equation}
\langle \hat{J}_{\alpha} \rangle = \langle \theta, \phi |\hat{R}%
^{\dagger}(0) e^{i\hat{H}^{\prime}} \hat{R}(t) \hat{J}_{\alpha} \hat{R}%
^{\dagger}(t) e^{-i\hat{H}^{\prime}} \hat{R}(0) | \theta, \phi \rangle . \label{J_alpha}
\end{equation}
In the first instance, this may be calculated by using the
commutation relation amongst the spin-$J$ components, namely $[\hat{J}_z,
\hat{J}_{\pm}] = \pm \hat{J}_{\pm}$ and $[\hat{J}_{+}, \hat{J}_{-}] = 2 \hat{%
J}_{z}$, and employing the Baker-Hausdorff theorem\cite{barnettradmore}. Useful identities are given in Eqs. (\ref{RJR}-\ref{RJmR}) with $\varphi(t) \equiv \Delta t$:
\begin{eqnarray}
\hat{R} \hat{J}_{z} \hat{R}^{\dagger} & = & - \frac{1}{2} ( \hat{J}_{+}e^{i
\Delta t} + \hat{J}_{-}e^{-i\Delta t} ),  \label{RJR} \\
\hat{R} \hat{J}_{+} \hat{R}^{\dagger} & = &  \frac{1}{2} ( \hat{J}_{+} -
\hat{J}_{-}e^{-2i\Delta t}) + \hat{J}_{z}e^{-i\Delta t}  ,  \label{RJpR}
\\
\hat{R} \hat{J}_{-} \hat{R}^{\dagger} & = &  \frac{1}{2} ( \hat{J}_{-} -
\hat{J}_{+}e^{2i\Delta t}) + \hat{J}_{z}e^{i\Delta t} .  \label{RJmR}
\end{eqnarray}
In principle, it is possible to repeatedly apply the Baker-Hausdorff theorem to obtain a general expression for $\hat{J}_{\alpha}(t) \equiv \hat{U}%
^{\dagger}(t)\hat{J}_{\alpha}\hat{U}(t)$; for the special case of $q=0$, the expressions may, in fact, be simplified further, giving
\begin{equation}
\langle \hat{J}_{z}(t) \rangle = \frac{N}{2} \cos \theta \cos [2\Omega
t+\Delta t + \phi] ,  \label{evJztq0}
\end{equation}
and
\begin{eqnarray}
\langle \hat{J}_{\pm}(t) \rangle &=& \frac{N}{2} e^{\mp i\Delta t} \left [
\cos \left(\frac{\theta - \pi/2}{2} \right ) \right.  \nonumber \\
 && \left. \mp i \sin \left(\frac{\theta - \pi/2}{2} \right ) \sin(2\Omega t + \Delta t + \phi) \right ]. \label{evJpmtq0}
\end{eqnarray}
It is found however that the expressions quickly become unwieldy especially for $q \neq 0$. The way around the problem is to note an important and useful point that, for the Hamiltonian under consideration, the transformation operator $\hat{R}$ happens to have the same form as the generator of $SU(2)$ atomic coherent states[Eq. (\ref{coherent})] if one writes  $\varphi(t) = -\phi$. In
particular, the effect of $\hat{R}$ is to rotate the CSS such that $\hat{R}%
(\theta ^{\prime },\phi ^{\prime })|\theta ,\phi \rangle \rightarrow |\theta
+\theta ^{\prime },\phi - \phi ^{\prime }\rangle $, i.e.
\begin{equation}
\hat{R}(t)|\theta ,\phi \rangle  = 
\sum_{m=-j}^{j}  {\cal R}_{m}^{j}(\theta - \pi/2, \phi  - \Delta t) |j,m\rangle ,  \label{Rphi}
\end{equation}
where ${\cal R}_{m}^{j}(\theta, \phi)$ is defined in Eq. (\ref{Rjmtp}).
Using the well-known relation for the raising and lowering operators $\hat{J}_{\pm}$, and the fact that $F(\hat{J}_{z})|j,m\rangle =F(m)|j,m\rangle $ where $F$ denotes some analytic function, one can obtain, after some algebra, an analytical expression for $\langle \hat{J}_{z}(t) \rangle$ as a summation:
\begin{eqnarray}
\langle \hat{J}_{z}(t)\rangle &=& -\sum_{m=-N/2}^{N/2-1} {\cal D}%
_{1}(\theta,m) \nonumber \\
&& \times \cos \left[2\Omega t - q\left( m+ \frac{1}{2}\right) t +
\Delta t + \phi \right],  \label{Jzt2}
\end{eqnarray}
where we have defined
\begin{eqnarray}
{\cal D}_{\alpha}(\theta,m) & = & C_{N/2+m+1}^{N}\left( \frac{N}{2}+m+1
\right )  \nonumber \\
&& \times \cos ^{2N}\left( \frac{\theta -\pi /2}{2} \right) \nonumber \\
&& \times  \tan^{N-2m-\alpha}\left( \frac{\theta -\pi /2}{2}\right).
\end{eqnarray}
Following somewhat more involved but identical steps as above, we obtain:
\begin{eqnarray}
\langle \hat{J}_{\pm}(t)\rangle &=& \sum_{m=-N/2}^{N/2-1} e^{\mp i (\Delta
t - \pi/2)} {\cal D}_{1}(\theta,m) \nonumber \\
&& \times \sin \left [ 2\Omega t - q \left ( m + \frac{1}{2} \right ) t + \Delta t + \phi \right ]  \nonumber \\
&& + \sum_{m=-N/2}^{N/2} e^{ \mp i \Delta t} {\cal D}_{0}(\theta,m) \left
( \frac{m}{j-m} \right ).  \label{Jpt2}
\end{eqnarray}
from which $\langle \hat{J}_{x}(t) \rangle$ and $\langle \hat{J}_{y}(t)
\rangle$ may be deduced: $\hat{J}_{x} = \frac{1}{2}( \hat{J}_{+} + \hat{J}_{-})$ and $\hat{J}_{y} = \frac{1}{2i}( \hat{J}_{+} - \hat{J}_{-})$. 

\begin{figure}
\begin{center}
\centerline{\includegraphics[height=7cm]{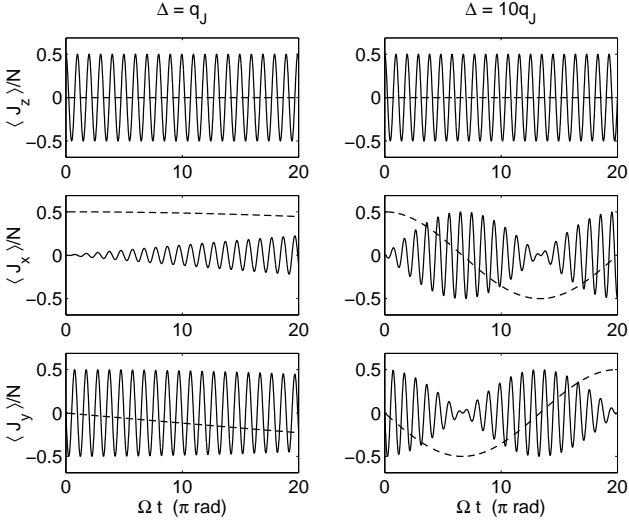}}
\caption{$\langle \hat{J}_{z}(t)\rangle/N$,  $\langle \hat{J}_{x}(t)\rangle/N$, and $\langle \hat{J}_{y}(t)\rangle/N$
for $q = 0$. Solid line: initial Dicke state, $|\theta = 0, \phi = 0 \rangle$ Dashed line: initial phase state $|\theta = \pi/2, \phi = 0 \rangle$. Left column: $\Delta = q_{j}$. Right column: $\Delta = 10q_{j}$, where $q_{j}/\Omega = 3/N$} \label{Jq0}
\end{center}
\end{figure}

\begin{figure}
\begin{center}
\centerline{\includegraphics[height=7cm]{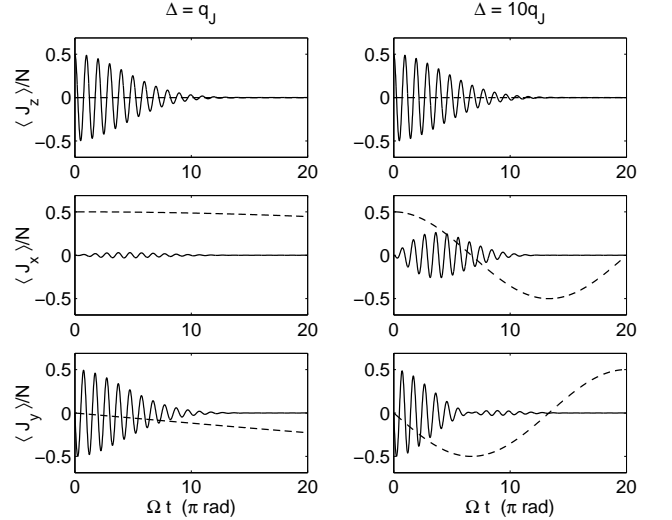}}
\caption{$\langle \hat{J}_{z}(t)\rangle/N$,  $\langle \hat{J}_{x}(t)\rangle/N$, and $\langle \hat{J}_{y}(t)\rangle/N$ for $q = q_{j}$. Identical details as in Fig. \ref{Jq0}.} \label{Jqsmall}
\end{center}
\end{figure}

We first present in Figs. \ref{Jq0} and \ref{Jqsmall}, the expectation
values $\langle \hat{J}_{\alpha}(t)\rangle$, $\alpha = x,y,z$ scaled to the total number of atoms $N$ for $q = 0$, and $q = q_{j}$ respectively for the detunings $\Delta = q_{j}$ and $\Delta = 10q_{j}$. It is noted that no significant qualitative difference is observed in these scaled amplitudes when one uses different number of atoms. In all the figures in this paper, the solid line represents the result for the initial CSS $|\theta, \phi=0 \rangle$ with $\theta = 0$ i.e. the Dicke state, while the dashed line corresponds to the initial phase state,  $\theta = \pi/2$. In Fig. \ref{Jq0}, it is seen that, for the initial state $|\theta = 0, \phi = 0 \rangle$, the $y$ and $z$ components of spin, $\langle \hat{J}_{y} \rangle$ and $\langle \hat{J}_{z} \rangle$, undergo oscillatory evolution with a $\pi/2$ phase shift. This implies that the macroscopic spin vector
may be visualized as if undergoing a circular motion in the $y$-$z$ plane. Superposed on this motion is the gradual increase of the oscillation in
the $x$ component, $\langle \hat{J}_{x} \rangle$ with the amplitude of the $y$ component being proportionally reduced with a $\pi/2$ phase shift. This implies that the spin vector initially undergoing a circular motion in the $y$-$z$ plane rotates around the $z$-axis with increasing amplitude.  This motion gets more pronounced for the larger $\Delta$ in that the spin vector undergoing circular motion in the $y$-$z$ plane rotates around the $z$-axis relatively quickly into a circular motion in the $x$-$z$ plane. This continues on to return to the $y$-$z$ plane, repeating this pattern over time.  For the initial state $|\theta = \pi/2, \phi = 0 \rangle$, there is no oscillation in the $y$-$z$ plane, and only a slow rotation through the $x$-$y$ plane is observed. This state may be identified as a ``self-trapping''
state, as there is no transfer of populations during the evolution. With higher $\Delta$, the frequency of this rotation is clearly increased.

\begin{figure}
\begin{center}
\centerline{\includegraphics[height=7cm]{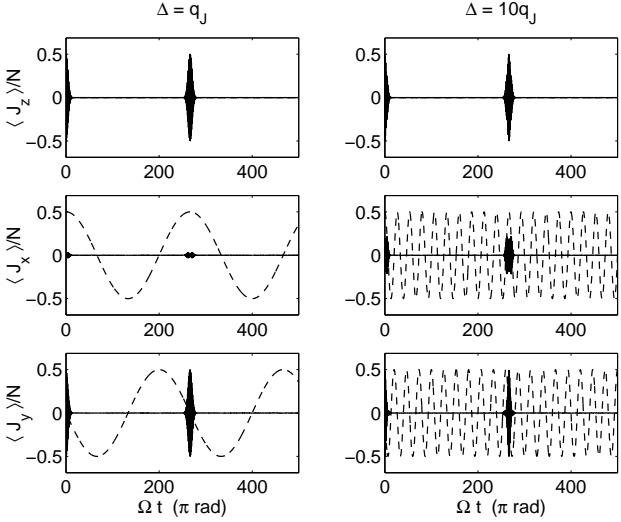}}
\caption{$\langle \hat{J}_{z}(t)\rangle/N$,  $\langle \hat{J}_{x}(t)\rangle/N$, and $\langle \hat{J}_{y}(t)\rangle/N$ for $q = q_{j}$, simulated for a longer time to observe revivals. Identical details as in Fig. \ref{Jq0}.} \label{Jqsmalllong}
\end{center}
\end{figure}

For the case of $q = q_{j}$, Fig. \ref{Jqsmall} shows the dephasing
or ``collapse'' of oscillations for the initial state $|\theta = 0, \phi = 0 \rangle$. 
The three spin components collapse to give average spin of almost zero. As will be seen below, this has a significant impact on the variance, and consequently on squeezing and entanglement in a TBEC. For the self-trapping initial state $|\theta = \pi/2, \phi = 0 \rangle$,
there is no change in the behavior with $q = q_{j}$. As to be expected, higher values of nonlinearity $q$ results in
shorter time scales for the collapse and with higher number of atoms, no significant qualitative difference is observed except that the collapse happens somewhat faster. The long time simulation is provided in Fig. \ref{Jqsmalllong}, demonstrating clear revivals.

The zero mean spin due to collapse indicates that the quantum state of a TBEC evolves from a CSS which is similar to the usual coherent state in quantum optics to a state which is similar to the number eigenstate with an equal number of atoms in each mode. The state that the CSS evolves into is, however, not exactly a self-trapping state as it clearly revives back into a CSS; it may be viewed as a quantum state with a small spread around the mean self-trapping state. For lack of better terminology we shall refer to this collapsed quantum state as a ``{\em quasi}-self-trapping state'' in this paper.

\subsection{Quantum squeezing and fluctuations}

In the language of spin squeezing, any reduction of the quantum noise for $q = 0$ corresponds to an effective squeezing due to the rotation of the coordinate axes, while squeezing observed for $q = q_{j}$ comes from a complex combination of coordinate rotation on top of the interatomic collision effect of the one-axis twisting term proportional to $\hat{J}^{2}_{z}$\cite{ueda}. 
In order to calculate the amount of squeezing in the system, we first calculate
the variances in the three components of the macroscopic spin vector, i.e. $%
[\Delta \hat{J}_{\alpha}(t)]^2 = \langle \hat{J}^{2}_{\alpha}(t) \rangle -
\langle \hat{J}_{\alpha}(t) \rangle^{2}$, $\alpha = x,y,z$ which, as mentioned above are 
experimentally accessible quantities. The unitary property of $\hat{R}$ i.e. $\hat{%
R}^{\dagger}\hat{R} = \openone$ is used to evaluate terms of the form $\hat{R}\hat{J}%
_{\alpha} \hat{J}_{\beta} \hat{R}^{\dagger} \equiv \hat{R} \hat{J}_{\alpha}
\hat{R}^{\dagger} \hat{R} \hat{J}_{\beta} \hat{R}^{\dagger}$ along with the identities Eqs. (\ref{RJR}-\ref{RJmR}). 

It can be shown that $\langle \hat{J}_{z}^{2}(t)\rangle$ is given by
\begin{eqnarray}
\langle \hat{J}_{z}^{2}(t)\rangle &=& \frac{1}{2} \sum_{m=-N/2}^{N/2-2}
{\cal D}_{2}(\theta,m) \left( \frac{N}{2} - m - 1 \right) \nonumber \\ 
& & \times \cos [4\Omega t - 2q(m+1) t + 2\Delta t + 2\phi]  \nonumber \\
& & + \frac{1}{4} \sum_{m=-N/2}^{N/2-1} {\cal D}_{2}(\theta,m) \left( \frac{N}{2} - m \right ) \nonumber \\
&& + {\cal D}_{0}(\theta,m) \left( \frac{N}{2} + m + 1
\right),  \label{Jzsqt2}
\end{eqnarray}
so that the variance in the $z$ direction, $\lbrack \Delta \hat{J}_{z}(t)]^{2}$, is
simply given by subtracting the square of Eq. (\ref{Jzt2}) from Eq. (\ref
{Jzsqt2}). On the other hand, $\langle \hat{J}^{2}_{x} \rangle = \frac{1}{4} [ \langle \hat{J}^{2}_{+} \rangle + \langle
\hat{J}^{2}_{-} \rangle + \langle \hat{J}_{+} \hat{J}_{-} \rangle + \langle
\hat{J}_{-} \hat{J}_{+} \rangle ]$ and $\langle \hat{J}^{2}_{y} \rangle =
\frac{1}{4} [\langle \hat{J}_{+} \hat{J}_{-} \rangle + \langle \hat{J}_{-} \hat{%
J}_{+} \rangle - \langle \hat{J}^{2}_{-} \rangle - \langle \hat{J}^{2}_{+}
\rangle]$.  The required expectation values of $\langle \hat{J}^{2}_{\pm} \rangle$ and $\langle \hat{J}_{\pm}\hat{J}_{\mp} \rangle$  are given by:
\begin{widetext}
\begin{eqnarray}
\langle \hat{J}_{+}^{2}(t)\rangle &=&  \sum_{m=-N/2}^{N/2-2}
\frac{e^{-2i \Delta t}}{2} {\cal D}_{2}(\theta,m) \left( \frac{N}{2} - m - 1 \right
) \cos [ 4\Omega t - 2q(m+1) t + 2\Delta t + 2\phi ]  \nonumber \\
&& - \sum_{m=-N/2}^{N/2-1}   \frac{e^{-2i \Delta t}}{4} \left \{
{\cal D}_{2}(\theta,m) \left( \frac{N}{2} - m \right ) +  {\cal D}_{0}(\theta,m) \left( \frac{N}{2} + m + 1 \right ) \right.  \nonumber \\ 
& & + \left. (2m+1)i  \sin \left [ 2\Omega t - q \left (m + \frac{1}{2} \right ) t + \Delta t + \phi \right ] \right \}  + \sum_{m=-N/2}^{N/2} e^{-2i \Delta t} {\cal D}_{0}(\theta,m) \left (\frac{m^2}{j-m} \right ),  \label{Jpsqt2}
\end{eqnarray}
\begin{eqnarray}
\langle \hat{J}_{+}(t)\hat{J}_{-}(t)\rangle &=& -
\sum_{m=-N/2}^{N/2-2} \frac{1}{2} {\cal D}_{2}(\theta,m) \left( \frac{N}{2} - m - 1
\right ) \cos[4\Omega t - 2q(m+1) t + 2 \Delta t + 2 \phi ]  \nonumber \\
&& +  \sum_{m=-N/2}^{N/2-1} \frac{1}{4} \left \{ {\cal D}_{2}(\theta,m) \left( \frac{N}{2} - m \right )  + {\cal D}_{0}(\theta,m) \left ( \frac{N}{2} + m + 1 \right ) \right.  \nonumber \\
&&  - \left. 4{\cal D}_{1}(\theta,m)  \cos \left [ 2\Omega t - q\left ( m + \frac{1}{2} \right ) t + \Delta t + \phi \right ] \right \} + \sum_{m=-N/2}^{N/2} {\cal D}_{0}(\theta,m) \left ( \frac{m^2}{j-m} \right ),  \label{Jpmsqt2}
\end{eqnarray}
\end{widetext}
with $\langle \hat{J}_{-}^{2}(t)\rangle \equiv \langle \hat{J}_{+}^{2}(t)\rangle^{*}$
where asterisk (*) denotes complex conjugate, and
$\langle \hat{J}_{-}(t)\hat{J}_{+}(t)\rangle =  \langle \hat{J}_{+}(t)\hat{J}_{-}(t)\rangle - 2  \langle \hat{J}_{z}(t) \rangle$, as to be expected from the commutation relation $[ \hat{J}_{+}, \hat{J}_{-}] = 2\hat{J}_{z}$. 
From these expressions and those for $\langle \hat{J}_{x}(t) \rangle$ and $\langle \hat{J}_{y}(t) \rangle$ obtained above, the variances  $[\Delta \hat{J}_{x}(t)]^{2}$ and  $\lbrack \Delta \hat{J}_{y}(t)]^{2}$ may be calculated.

\begin{figure}
\begin{center}
\centerline{\includegraphics[height=7cm]{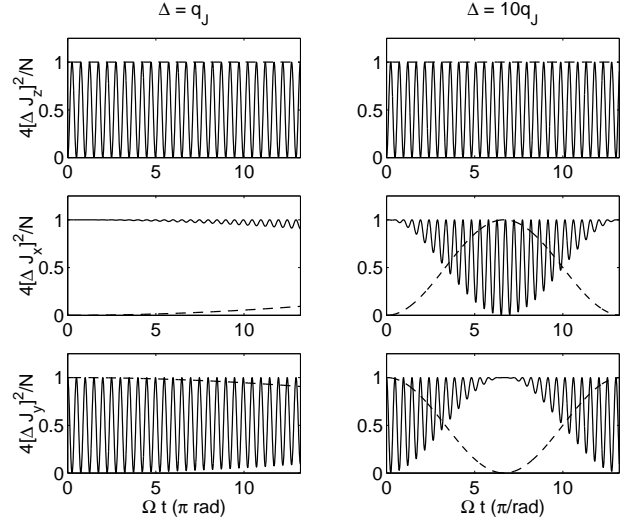}}
\caption{Variances $[\Delta \hat{J}_{z}(t)]^2$,  $[\Delta \hat{J}_{x}(t)]^2$, and $[\Delta \hat{J}_{y}(t)]^2$ for $q = 0$, scaled to SQL of $N/4$ for clarity. Scaled variance of 1 in this figure corresponds to the SQL. Identical details as in Fig. \ref{Jq0}.} \label{Varq0}
\end{center}
\end{figure}

We find that, for $q=0$, the variances $[\Delta \hat{J}%
_{\alpha}]^2$, $\alpha = x,y,z$, show an oscillatory behavior which is
bounded above by $N/4$ [Fig. \ref{Varq0}]. This is an expected result
for a CSS, which is known to have variances in the standard quantum limit
(SQL) of $J/2$. One key observation regarding Fig. \ref{Varq0} is
that the variances do go below the SQL. This kind of squeezing which is due
to the rotation of the coordinate axis\cite{ueda} naturally occurs since a
spin vector is an eigenstate of spin in one direction with zero variance in that direction. 
As the spin vector
traverses the phase space due to dynamical evolution, the error ellipsoid
follows the path of the spin vector in such a way that the minor axis of the
ellipsoid periodically lines up with the $x$, $y$ or $z$ axis, resulting in the 
reduction of quantum fluctuations in that direction.
For small $\Delta$, the variance in the $z$ direction show oscillatory reduction with $\pi/2$ phase shift with that of the $y$ component. The variance in the $x$ direction is seen as more or less maintaining its value near the SQL. For larger $\Delta$, the uncertainty in the $x$ direction is found oscillating with increasing amplitude, which is consistent with the error ellipsoid following the motion of the rotating spin vector discussed above.  

\begin{figure}
\begin{center}
\centerline{\includegraphics[height=7cm]{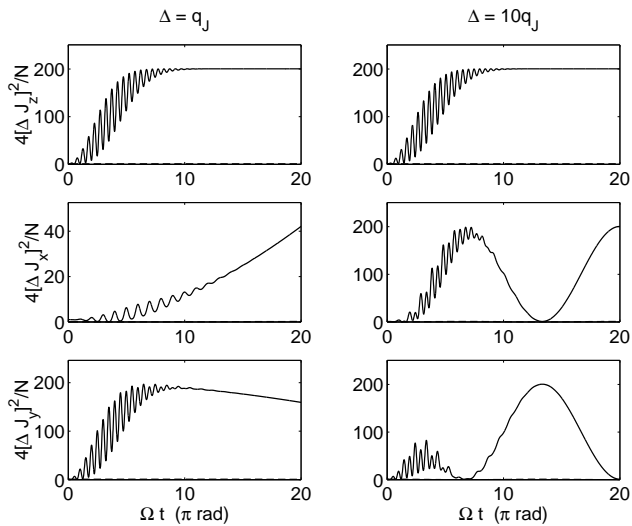}}
\caption{Variances $[\Delta \hat{J}_{z}(t)]^2$,  $[\Delta \hat{J}_{x}(t)]^2$, and $[\Delta \hat{J}_{y}(t)]^2$ for $q = q_{j}$, again scaled to the SQL, $N/4$. Identical details as in Fig. \ref{Jq0}.} \label{Varqsmall}
\end{center}
\end{figure}

\begin{figure}
\begin{center}
\centerline{\includegraphics[height=7cm]{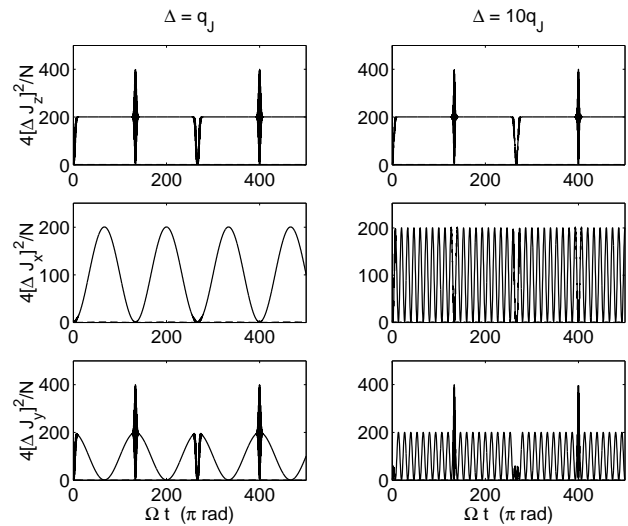}}
\caption{Variances $[\Delta \hat{J}_{z}(t)]^2$,  $[\Delta \hat{J}_{x}(t)]^2$, and $[\Delta \hat{J}_{y}(t)]^2$ for $q = q_{j}$ scaled to the SQL, $N/4$, simulated for a longer time to observe revivals. Identical details as in Fig. \ref{Jq0}.} \label{Varqsmalllong}
\end{center}
\end{figure}

For the $q = q_{j}$ case, the variances in the $x$, $y$, and $z$
directions give significantly different behavior, as shown in Fig. \ref{Varqsmall}.  The unexpected feature is the large variance equivalent to the maximum relative uncertainty (standard deviation) of order $\pm 35 \%$ per measurement. Mathematically, this can be understood as the corollary of the collapsing mean spin components; the collapsing spin vector implies that the the variance $[\Delta \hat{J}_{\alpha}]^2 = \langle \hat{J}_{\alpha}^2 \rangle - \langle \hat{J}_{\alpha} \rangle^2$ is dominated by the $\langle \hat{J}_{\alpha}^2 \rangle$ term of the order $N^2$ as it cannot be cancelled by the $\langle \hat{J}_{\alpha} \rangle^2$ term. Careful analysis of the variance for the $q=0$ case discussed above reveals that the terms of the form $\langle \hat{J}_{\alpha}^2 \rangle$ are almost exactly cancelled by the $\langle \hat{J}_{\alpha} \rangle^2$ which are also of the order $N^2$ to maintain the variances of the order of the SQL, $N/4$. This type of cancellation cannot happen as the spin vector collapses with $q = q_{j}$.  Physically, the increased uncertainty can be attributed to the quantum state evolving away from the minimum uncertainty state of CSS into a quasi-self-trapping state with much higher quantum fluctuations. In order to display periodic behavior in the variances, we plot in Fig. \ref{Varqsmalllong} the variances over the identical time period as in Fig. \ref{Jqsmalllong}. The result is in agreement with the purely numerical result of Tonel {\em et al.}\cite{tonel}. It is found that, as the revivals occur, the variances do go below the SQL periodically.  Although it is not clearly depicted due to scaling, the results for the initial phase state $| \theta = \pi/2, \phi = 0 \rangle$ are identical to that for the $q = 0$ case i.e. the variance stays at 1 for the atom number difference, and oscillates between $0$ and the SQL for the relative phase.


\subsection{Dynamical evolution of quantum entanglement in a TBEC}

As mentioned in Ref. \cite{Milburn}, the amount of entanglement depends on the way a system is partitioned into subsystems. In this subsection we calculate the dynamical evolution of quantum entanglement in two different subsystem decomposition of the TBEC: entanglement between the particles and that between the two modes. 

\subsubsection{Entanglement between the particles}

The atom-atom entanglement is parameterized by Eq. (\ref{Sxi}). Equation (\ref{Sxi}) is useful since, regardless of the actual spin
direction at a given time, one need only to identify three unit orthogonal directions ${\bf n}_{i}$ to check whether the system has become entangled due to system dynamics. If one's goal is to identify maximum entanglement possible between the atoms, it is necessary to scan through all
possible ${\bf n}_{i}$'s within the unit sphere at each point in time. However, it should be noted that keeping to a
fixed direction may help in constructing an actual experimental scheme to measure $\xi_{\alpha}^{2}(t)$. In particular, we consider in this paper experimentally meaningful variances and amplitudes of the macroscopic spin vector in the $x$, $y$, and $z$ directions. 
We therefore consider
\begin{equation}
\xi^{2}_{\alpha}(t) =\frac{N[\Delta \hat{J}_{\alpha}(t)]^{2}}{%
\langle \hat{J}_{\beta}(t)\rangle ^{2}+ \langle \hat{J}_{\gamma}(t)\rangle ^{2}}
\end{equation}
where $\alpha, \beta, \gamma$ cycle through $x, y, z$. 

\begin{figure}
\begin{center}
\centerline{\includegraphics[height=7cm]{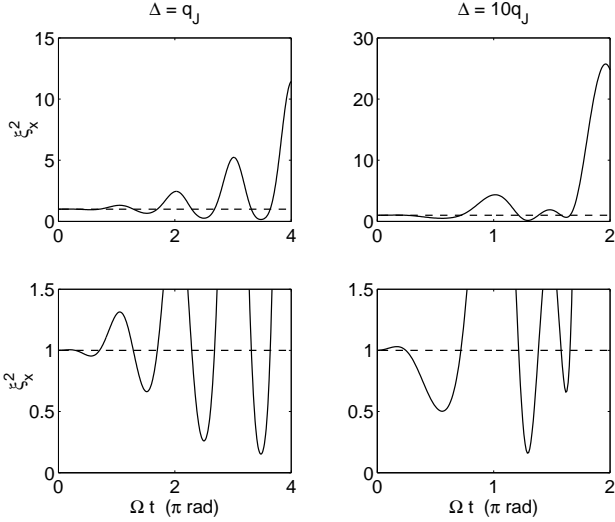}}
\caption{Entanglement parameter $\xi_{x}^2$
for $q = q_{j}$. Identical details as in Fig. \ref{Jq0}.} \label{Xiqsmall}
\end{center}
\end{figure}

For $q=0$, it was found that $\xi^{2}_{\alpha}(t) \equiv 1$, $\alpha = x,y,z$, at all times for both initial states i.e. not entangled despite the fact that, as already seen in Fig. \ref{Varq0}, squeezing due to the rotation of the coordinate
axis does occur. This shows that the observed squeezing is not accompanied by an increase in quantum correlations between the particles.  On the other hand, for $q=q_{j}$, it is found that $\xi^{2}_{\alpha}$ is maintained at 1 for the initial phase state as to be expected, while for the initial Dicke state, the very quickly increasing quantum fluctuations and the collapsing spin vector due to dephasing   $\xi_{\alpha}^{2}(t)$ large.  Physically, this is due to the fact that the CSS is evolving into a quasi-self-trapping quantum state with low atom-atom correlations. A correct balance between the rapidly increasing variance and the collapsing spin vector is clearly needed in order to maintain entanglement with $\xi_{\alpha}^2 (t) <1$.  From Fig. \ref{Varqsmall}, it is seen that the variance in $\hat{J}_{x}$ increases most slowly amongst the variances, while Fig. \ref{Jqsmall} indicates the spin vector $\langle \hat{J}_{y} \rangle + \langle \hat{J}_{z} \rangle$ oscillating with a non-zero amplitude, showing the most promise in finding entanglement in this spin direction.  In Fig. \ref{Xiqsmall} we show $\xi^2_{x} (t)$ for  $q = q_{j}$,  for $\Delta = q_j$ and $\Delta = 10q_j$. Differently from other results, it was found that with $\Delta = 0$, $\xi^2_{x}(t) \geq 1$ for all times, indicating the crucial role $\Delta$ plays in this system. For the initial Dicke state the parameter quickly becomes large, clearly indicating that quantum correlations between the atoms are destroyed rapidly. However, it is seen that the parameter $\xi_{x}^2 (t)$ does dip below 1 for a brief time period demonstrating atom-atom entanglement. 

\begin{figure}
\begin{center}
\centerline{\includegraphics[height=7cm]{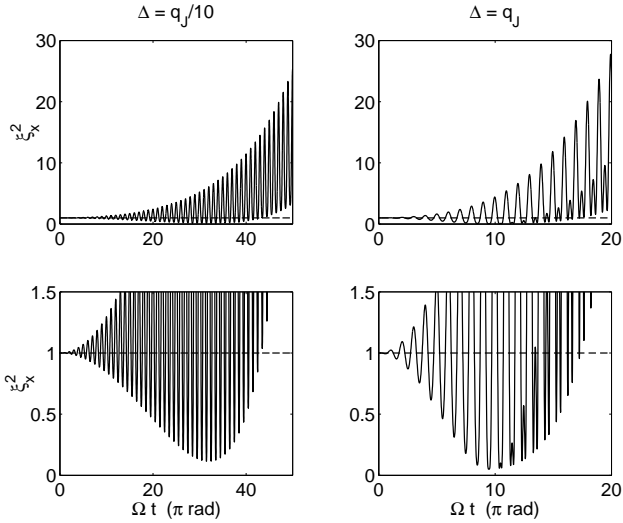}}
\caption{Entanglement parameter $\xi_{x}^2$
for reduced nonlinearity, $q = q_{j}/10$. Identical details as in Fig. \ref{Jq0}.} \label{Xiqsmaller}
\end{center}
\end{figure}

Somewhat counter-intuitively, one may extend the duration over which the atoms are entangled by {\em reducing} the nonlinearity constant $q$; it is clear that with a smaller $q$, the variance in $\hat{J}_{x}$ increases more gently while the collapse of the spin vector occurs over a longer time scale, extending the duration over which $\xi^2_{x} < 1$. This is presented in Fig. \ref{Xiqsmaller} for $q = q_{j}/10$. As before, the increase in $\Delta$ results in the increase in oscillation frequency. This indicates a potential for the quantum control of entanglement properties of a TBEC   via externally adjustable parameters.

\subsubsection{Entanglement between the two modes}

\begin{figure}
\begin{center}
\centerline{\includegraphics[height=7cm]{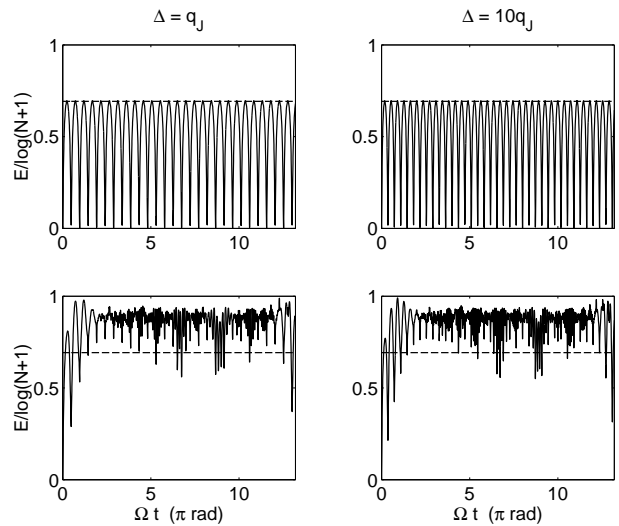}}
\caption{Scaled von Neumann entanglement parameters $E$ for $N = 40$. Top row:  $q = 0$, Bottom row: $q = q_{j}$. Left column: $\Delta = q_{j}$, Right column: $\Delta = 10q_{j}$} \label{E}
\end{center}
\end{figure}

Writing the quantum state of a TBEC in the form 
\begin{eqnarray}
|\Psi (t) \rangle & = & \sum_{m = 0}^{N} c_{m}(t) |N\rangle |N-m \rangle \\
& = & \sum_{m = -j}^{j} c_{m}(t) |j, m \rangle
\end{eqnarray}
the well-known von Neumann entropy that shows degree of entanglement can be written\cite{Milburn}:
\begin{equation}
E(t) = -\frac{1}{\log_2(N+1)} \sum_{m = -j}^{j} |c_m(t)|^2 \log_2 |c_m(t)|^2  \label{scaledE}
\end{equation}
where the normalization factor $\log_2(N+1)^{-1}$ was included so that $0< E < 1$. The expansion coefficients are given by
\begin{eqnarray}
c_m(t) & = & \langle m | \hat{U}(t) | \psi(0) \rangle \\
& = & \langle m | \hat{R}^{\dagger}(t) e^{-i H' t} \hat{R}(0) | \theta, \phi \rangle , 
\end{eqnarray} 
where the rotation operator $\hat{R}$, reduced Hamiltonian $H'$ and the CSS $|\theta, \phi \rangle$ are as defined above. Using Eq. (\ref{Rphi}), and  inserting the completeness relation $\sum_{m'} |m' \rangle\langle m'| = \openone$ one can write
\begin{eqnarray}
c_m(t) & = & \sum_{m'} {\cal F}^{(\lambda)}_{m,m'}(t) {\cal R}_{m'}^{j}(\theta - \pi/2,\phi -\Delta t)  \nonumber \\  
&& \times e^{-i[2m'\Omega t - qm'^2 t/2]},   \label{cmt}
\end{eqnarray} 
where 
\begin{eqnarray}
{\cal F}^{(\lambda)}_{m,m'}(t)  & = &  \langle m | \hat{R}^{\dagger}(t) |m' \rangle 
\end{eqnarray}
represents the matrix element of the rotation operator. This can be evaluated by first
applying the disentangling theorem on the rotation operator $\hat{R}$\cite{bloch}:
\begin{eqnarray}
\hat{R}(\lambda, \Delta t) & = & \exp \left [ \frac{\lambda}{2} ( \hat{J}_{-}e^{-i\Delta t} - \hat{J}_{+}e^{i\Delta t} ) \right ] \\ 
& = & e^{\tau \hat{J}_{+}}e^{\ln (1 + |\tau|^2 ) \hat{J}_{z}}e^{-\tau^{*} \hat{J}_{-}} ,
\end{eqnarray}
where $\tau  = e^{i\Delta t}\tan (-\frac{\lambda}{2})$. Taking into account the unitary nature of the rotation operator$\hat{R}^{\dagger}(\theta, \phi) = \hat{R}(-\theta, \phi)$ and expanding the exponential operators containing $\hat{J}_{+}$ and $\hat{J}_{-}$ one can write
\begin{eqnarray}
{\cal F}^{(\lambda)}_{m,m'}(t) & = &  \sum_{n, n'} \frac{(-\tau)^{n}}{n!} \frac{\tau^{* n'}}{n' !} (1+ |\tau|^2)^{m' - n' } \nonumber \\
&& \times \langle m | \hat{J}_{+}^{n} \hat{J}_{-}^{n' }|m' \rangle .
\end{eqnarray}
The final expression for the matrix element, which was obtained by using the ladder operator nature of $\hat{J}_{\pm}$ and found to be identical to the irreducible representation of full rotation group \cite{bloch} is:
\begin{eqnarray}
{\cal F}^{(\lambda)}_{m,m'}(t) & = &  \sum_{n = 0}^{2j} \frac{(-1)^{m'-m+n}}{(n + m' - m)! n!} \frac{(j - m + n)!}{(j + m - n)!}   \nonumber \\
&& \times   \left [ \frac{(j + m)!(j+m')!}{(j - m)!(j-m')!} \right ]^{1/2} e^{i(m-m')\Delta t}  
\nonumber \\
&& \times  \sin^{2n + m' - m} \left ( \frac{\lambda}{2} \right )\cos^{-m'-m} \left (\frac{\lambda}{2} \right ). \label{Fmatrix}
\end{eqnarray}
The normalized entanglement parameter $E(t)$ is plotted in Fig. \ref{E}, for the four combinations of $q$ and $\Delta$: $q = 0$, $q = q_{j}$ and $\Delta = q_{j}$, $\Delta = 10q_{j}$. It is seen that the nonlinearity is crucial for achieving high degree of entanglement. The effect of $\Delta$ was, as expected, to increase the frequency of oscillations. To generate Fig. \ref{E}, we used Eqs. (\ref{scaledE}), (\ref{cmt}), and (\ref{Fmatrix}) with $N=40$ due to the computational limitations in numerically calculating factorials of large numbers. However, it was found that, with non-zero $q$, as long as the ratio of the various parameters such as $q/N$ is kept the same, there are no discernible changes in the plot as a function of $N$. Our result agrees with the numerical result obtained by Tonel {\it et al.}\cite{tonel} with higher number of atoms.  

\begin{figure}
\begin{center}
\centerline{\includegraphics[height=5cm]{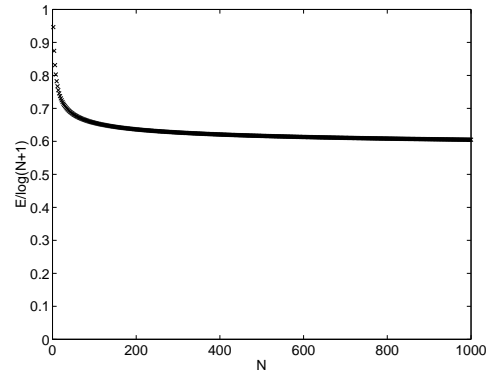}}
\caption{Scaled von Neumann entanglement parameters $E$ as a function of the number of atoms $N$ for the initial phase state.} \label{Ephase}
\end{center}
\end{figure}

It was found that, for the initial phase state $|\theta = \pi/2, \phi = 0 \rangle$, the entanglement parameter does depend on the number of atoms, as the normalized coefficients $|c_m(t)|^2$ are given in this case by the binomial distribution, namely, $|c_m(t)|^2 \equiv C^{2j}_{j+m}$. It is also notable that the coefficients are no longer time dependent. As the number of particles $N$ and hence $j$ increases, the broad binomial distribution approximates narrower and narrower Gaussian as a function of $m$, and hence the state becomes less and less entangled. For example, for $N = 400$, it is found that the scaled $E = 0.62$.  In Fig. \ref{Ephase}, we plot the entanglement parameter as a function of the number of atoms $N$ for the initial phase state. It is seen that the value tends towards 0.6 as the number is increased up to 1000 atoms. Such consistent behavior exhibited for the bipartite entanglement suggests a possibility to use this property to create a macroscopic matter-wave state with known entanglement.

\section{Conclusion}

We have studied the dynamical evolution of a TBEC in the presence of an adiabatically varying, off-resonant atom-light coupling.  The macroscopic,
fictitious spin vector was found to undergo a rather complex motion, and further complications in the form of collapses and revivals were found with nonzero nonlinearity. As the main result of this paper, we used an exact solution to the time-dependent Schr\"{o}dinger equation to calculate analytically the amount of quantum mechanical squeezing and entanglement in a TBEC under various conditions. In particular, we considered entanglement generated between the atoms and that between the modes. For the case with the nonlinearity turned off, although the variance was found to be reduced below the SQL (i.e. squeezed), the system never demonstrated atom-atom entanglement, and the entanglement between the two modes remained well below its maximal value.
With the nonlinearity turned on, it was found that the atom-atom entanglement can be generated initially, and then become unentangled rapidly.  On the other hand, the entanglement between the two modes was found to more or less maintain maximal values throughout, except for some fluctuations. The dynamics of entanglement was found to be controllable via various parameters present in this system, namely the nonlinearity $q$ and the detuning $\Delta$.

Potentially useful results of this work are the squeezing below the SQL in the atom number difference and relative phase which may help reduce the projection noise in spectroscopic and interferometric applications as discussed by Wineland {\it et al.}\cite{wineland}, and the consistently high degree of entanglement between the two modes with the nonlinearity turned on, which could be useful in the context of quantum information science. In addition, the dynamically stable features of the self-trapping state and the time-independent entanglement between the two modes which is also relatively insensitive to the changes in the atom number (for large $N$) could prove useful in designing a robust, macroscopic bipartite quantum state with known entanglement. Future work could involve studying quantum control methods for maintaining optimal squeezing and entanglement in a TBEC, for such applications as precision matter-wave interferometry using entangled BECs\cite{interferometry}.


\begin{thebibliography}{99}

\bibitem{squeeze1}  J. A. Dunningham, M. J. Collett, and D. F. Walls, Phys.
Lett. A {\bf 245} 49 (1998)

\bibitem{squeeze2}  J. Rogel-Salazar, S. Choi, G. H. C. New, and K. Burnett,
Phys. Lett. A {\bf 299} 476 (2002)

\bibitem{squeeze3}  J. Rogel-Salazar, G. H. C. New, S. Choi, and K. Burnett,
Phys. Rev. A {\bf 65} 023601 (2002)

\bibitem{squeeze4}  V. Chernyak, S. Choi, and S. Mukamel, Phys. Rev. A {\bf 67}, 053604
(2003)

\bibitem{squeeze5}  S. Choi, and N. P. Bigelow, J. Mod. Opt. {\bf 52}, 1081 (2005)

\bibitem{WallsNature}  D. F. Walls, Nature {\bf 306}, 141 (1983); D. F.
Walls and G. J. Milburn, {\it Quantum Optics} (Springer-Verlag, Berlin, 1994)



\bibitem{wallszoller}  D. F. Walls, and P. Zoller, Phys. Rev. Lett. {\bf 47}%
, 709 (1981)

\bibitem{ueda}  M. Kitagawa, and M. Ueda, Phys. Rev. A {\bf 47}, 5138 (1993)

\bibitem{wineland}  D. J. Wineland, J. J. Bollinger, W. M. Itano, and D. J.
Heinzen, Phys. Rev. A {\bf 50}, 67 (1994)

\bibitem{twocomp1}  A. S\/{o}rensen, L.-M. Duan, J. I. Cirac, and P. Zoller
Nature (London) {\bf 409}, 63 (2001)

\bibitem{Micheli}  A. Micheli, D. Jaksch, J. I. Cirac, and P. Zoller, Phys.
Rev. A, {\bf 67}, 013607 (2003)

\bibitem{kennedy}  S. D. Jenkins, and T. A. B. Kennedy Phys. Rev. A {\bf 66}%
, 043621 (2002)

\bibitem{twocomp2}  L.-M. Duan, A. S\/{o}rensen, J. I. Cirac, and P. Zoller
Phys. Rev. Lett. {\bf 85}, 3991 (2001)

\bibitem{twocomp3}  M. G. Moore and P. Meystre Phys. Rev. A {\bf 59}, R1754
(1999)

\bibitem{Milburn}  A. P. Hines, R. H. McKenzie, and G. J. Milburn Phys. Rev. A {\bf 67}, 013609 (2003)

\bibitem{myatt}  C. J. Myatt, E. A. Burt, R. W. Ghrist, E. A. Cornell, and C. E. Wieman,
Phys. Rev. Lett {\bf 78}, 586 (1997)

\bibitem{stenger}  J. Stenger, S. Inouye, D.M. Stamper-Kurn, H.-J. Miesner, A.P. Chikkatur, and W. Ketterle,
Nature (London) {\bf 396}, 345 (1998);
H.-J. Meisner, D.M. Stamper-Kurn, J. Stenger, S. Inouye, A.P. Chikkatur, and W. Ketterle,
 Phys. Rev. Lett. {\bf 82}, 2228 (1999)

\bibitem{bloch}  F. T. Arrechi, E. Courtens, R. Gilmore, and H. Thomas,
Phys. Rev. A {\bf 6}, 2211 (1972)

\bibitem{barnettradmore}  S. M. Barnett and P. M. Radmore, {\it Methods in
Theoretical Quantum Optics} (Oxford University Press, Oxford, 1997)

\bibitem{choi}  I. Fuentes-Guridi, J. Pachos, S. Bose, V. Vedral, and S.
Choi Phys. Rev. A {\bf 66}, 022102 (2002)

\bibitem{chen}  Z.-D. Chen, J.-Q. Liang, S.-Q. Shen, and W.-F. Xie Phys.
Rev. A {\bf 69}, 023611 (2004)

\bibitem{feshbach}  S. Inouye, M.R. Andrews, J. Stenger, H.-J. Miesner, D.M. Stamper-Kurn, and W. Ketterle,
Nature (London) {\bf 392}, 151 (1998)

\bibitem{liang}  J.-Q. Liang, and H. J. W. M\"{u}ller-Kirsten, Ann. Phys.
(N.Y.) {\bf 219}, 42 (1992); Y. Z. Lai, J.-Q. Liang, H. J. W.
M\"{u}ller-Kirsten, and J.-G. Zhou, Phys. Rev. A {\bf 53}, 3691 (1996)

\bibitem{tunneling}  L.-M. Kuang, and Z.-W. Ouyang, Phys. Rev. A {\bf 61},
023604 (2000); W.-D. Li, X. J. Zhou, Y. Q. Wang, J. Q. Liang, and W. M. Liu,
Phys. Rev. A {\bf 64}, 015602 (2001)

\bibitem{savage}  D. Gordon, and C. M. Savage, Phys. Rev. A {\bf 61}, 023604 (1999)

\bibitem{Castin} Y. Castin and J. Dalibard, Phys. Rev. A {\bf 55}, 4330 (1997)

\bibitem{leggett} A. J. Leggett, Rev. Mod. Phys. {\bf 73}, 307 (2001)

\bibitem{tonel}  A. P. Tonel, J. Links, and A. Foerster, J. Phys. A: Math. Gen. {\bf 38}, 1235 (2005)

\bibitem{interferometry}  J. A. Dunningham, K. Burnett, and S. M. Barnett, Phys. Rev. Lett. {\bf 89}, 150401 (2002)

\end{thebibliography}
\end{document}